\documentclass[twocolumn,showpacs,preprintnumbers,amsmath,amssymb]{revtex4}
\usepackage{graphicx}% Include figure files
\usepackage{dcolumn}% Align table columns on decimal point
\usepackage{bm}% bold math
\usepackage{epsfig}
\usepackage{CJK}
\usepackage{bm}

\begin{document}

%\begin{CJK*}{GBK}{song} % Use default fonts from CJK (see below)
\title{Time-Reversal-Symmetry-Broken Superconductivity Induced by
Frustrated Inter-Component Couplings}

\author{Xiao Hu and Zhi Wang}

\affiliation{WPI Center for Materials Nanoarchitectonics,
National Institute for Materials Science, Tsukuba 305-0044, Japan}

\date{\today}

\begin{abstract}
Superconductivity is associated with spontaneously broken gauge
symmetry. In some exotic superconductors the time-reversal symmetry
is broken as well, accompanied with internal magnetic field. A
time-reversal symmetry broken (TRSB) superconductivity without
internal magnetic field involved can be induced by frustrated
inter-component couplings, which becomes a realistic issue recently
due to the discovery of iron-pnictide superconductors. Here we
derive stability condition for this novel TRSB state using the
Ginzburg-Landau (GL) theory. We find that there are multiple
divergent length scales, and that this novel superconductivity
cannot be categorized by the GL number into type I or type II. We
reveal that the critical Josephson current of a constriction
junction between two bulk superconductors of different chiralities
is suppressed significantly from that for same chirality. This
effect provides a unique way to verify experimentally this brand new
superconductivity.
\end{abstract}


 \maketitle
%\end{CJK*}

Spontaneous symmetry breaking is one of the most fundamental
concepts of modern physics, which plays a central role in particle
physics, condensed matter physics, and even in our understanding
into the birth of universe. Broken gauge symmetry is known as the
hallmark of superconductivity \cite{BCS, Nambu}. Simultaneous
breaking of gauge symmetry and time-reversal symmetry has
attracted considerable attentions \cite{SigristUeda}, with the
triplet superconductivity in Sr$_2$RuO$_4$ as an icon \cite{Maeno}.

Possibility of time-reversal-symmetry-broken (TRSB) superconductivity
in multi-component system was
discussed before in the context of conventional mechanism for exotic
superconductivity \cite{Gorkov}. Interests in this novel phenomenon
are renewed \cite{Tesanovic, Yanagisawa} by the discovery of
iron-based superconductors \cite{Hosono}. Previous works revealed
that frustrated inter-component couplings induce intrinsically complex
order parameters, and thus the TRSB superconductivity \cite{Gorkov,
Tesanovic, Yanagisawa}, primarily based on a special situation with all
components equivalent (the \textit{isotropic} system).

It was also discussed that a junction structure between
superconductors of two components and single component can exhibit a
similar TRSB situation \cite{NgNagaosa}. Possibility of TRSB state
was discussed in iron-based superconductors with coexisting s- and
d-wave order parameters \cite{SCZhang}. A novel dynamic mode was
proposed for three-component, time-reversal symmetry reserved (TRSR)
superconductivity \cite{Ota}. All these make the physics of
multi-component superconductivity very rich.

In the present work, we consider an \textit{anisotropic} system with
frustrated inter-component couplings. For simplicity, each component
is considered as s-wave and well described by the BCS theory
\cite{BCS}. We adopt the GL approach \cite{Tinkham}, powerful for
discussions on stability of states and thus phase transition,
symmetry breaking, as well as magnetic responses of
superconductivity, which are the primary concerns of the present
study. Benefitting from the simplicity of the GL theory, the
condition for stable TRSB state are derived explicitly. It is also
found that there are multiple divergent length scales, and that this
novel superconductivity cannot be categorized by the GL number into
type I and type II. We reveal that the critical Josephson current of
a constriction junction between two bulk superconductors of
different chiralities (see Fig.~1) is suppressed significantly from
that for same chilarity. A standard Josephson junction measurement
can provide smoking gun evidence for this brand new
superconductivity.

\begin{figure}[t]
\psfig{figure=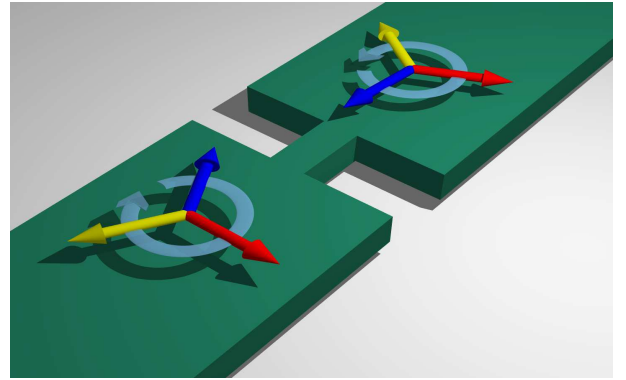,width=8cm}
 \caption{
TRSB states characterized by complex order parameters and opposite
chiralities of a three-component superconductor in a Josephson
junction of constriction structure.}
\end{figure}

The density of GL free-energy functional for a multi-component
superconductor is:

\vspace{-3mm}
\begin{equation}
\begin{array}{l}
 f=\sum\limits_{j}
    {\left[a_j|\psi_j|^2+\frac12 b_j|\psi_j|^4
     +\frac{1}{2m_j}\left|\left(\frac{\hbar}{i}\nabla-\frac{2e}{c}\mathbf{A}\right)\psi_j\right|^2
    \right]} \\\rm{\ \ \ \ \ \ \ \ \ }
 -\sum\limits_{j<k}\gamma_{jk}(\psi^{}_j\psi^*_k+c.c.) + \frac{1}{8\pi}(\nabla \times
 \mathbf{A})^2,
\end{array}
\label{GL}
\end{equation}

\noindent where summations running over the number of components are
understood, and all quantities are conventionally defined (see
\cite{Tinkham, Gorkov, Gurevich, Yanagisawa}); $a_j(T_{{\rm c}j})=0$
specify the transition points for individual components. To be
specific, we discuss here a system with three components, and
extension of the following discussions to more components is
straightforward. It is noticed that the physics addressed below
remains unchanged even when other possible terms (up to the quartic
order of order parameters) are included in Eq.(\ref{GL}).

For $\gamma_{12}\gamma_{23}\gamma_{13}>0$, the system behaves
basically as a single-band superconductivity close to the critical
point of the composite system. Hereafter, we focus on the case
$\gamma_{12}\gamma_{23}\gamma_{13}<0$, and treat a
system with all $\gamma_{jk}$'s negative, noticing that a simple
gauge transformation in GL free energy (\ref{GL}) links it to other
possible cases. The situation treated here may be realized in
over-doped iron pnictides \cite{Tesanovic}.

Around the critical point, the GL equations in absence of magnetic
field can be linearized as

\vspace{-3mm}
\begin{equation}
\left[
\begin{array}{ccc}
  a_1 & -\gamma_{12} & -\gamma_{13} \\
  -\gamma_{12} & a_2 & -\gamma_{23} \\
  -\gamma_{13} & -\gamma_{23} & a_3
\end{array}
\right]
\left[
\begin{array}{c}
  \psi_1 \\
  \psi_2 \\
  \psi_3
\end{array}
\right]
=\left[
\begin{array}{c}
  0 \\
  0 \\
  0
\end{array}
\right],
 \label{LGL}
\end{equation}

\noindent or in a vector form ${\bf Q}\cdot \Psi={\bf 0}$ with
coupling matrix ${\bf Q}$. The critical point of the composite
superconductivity $T_{\rm c}$ is given by the highest temperature
where the determinant of ${\bf Q}$ becomes zero, i.e.

\vspace{-3mm}
\begin{equation}
 a_1 a_2 a_3-2 \gamma_{12}\gamma_{23}\gamma_{13}
 -a_1\gamma^2_{23}-a_2\gamma^2_{13}-a_3\gamma^2_{12}=0,
\label{Tc}
\end{equation}

\noindent where temperature dependence of the quantities is
understood (see Fig.~2). From Sylvester's criterion \cite{ArfkenWeber}, one has
$a_{j}>0$ and $a_ja_k-\gamma^2_{jk}\geq 0$ at $T_{\rm c}$, namely
the critical point of the composite superconductivity is above any
single components, and not below any of the two-component ones, in
spite of negative couplings \cite{Kondo}.

If Eq.(\ref{Tc}) has a single root at $T=T_{\rm c}$ (see Fig.~2), or
equivalently there are two independent vectors in the coupling
matrix ${\bf Q}$, the ratios among the order parameters given by the
Cramer's rules for the components of the matrix ${\bf Q}$ should be
real \cite{ArfkenWeber}. The order parameters of three-component
superconductivity that minimize the GL free energy (\ref{GL}) can
then be taken as real numbers, apart from a common phase factor,
same as single- and two-component supernconductivity.

Equation (\ref{Tc}) has a doubly degenerated root \cite{note1} (see
Fig.~2), or equivalently there is only one independent vector in
${\bf Q}$, when

\vspace{-3mm}
\begin{equation}
  a_1 a_2-\gamma^2_{12}=0, \hspace{3mm}
  a_1 a_3-\gamma^2_{13}=0, \hspace{3mm}
  a_2 a_3-\gamma^2_{23}=0,
\label{condition1}
\end{equation}

\noindent at $T_{\rm c}$, namely $T_{\rm c}=T_{\rm c12}=T_{\rm
c13}=T_{\rm c23}$. The single independent vector in the coupling
matrix ${\bf Q}$ leaves the room for complex order parameters in
Eq.(\ref{LGL}) despite that all the parameters in ${\bf Q}$ are
real. Equation (\ref{condition1}) is the first condition for a TRSB
state specified by complex order parameters.

It is easy to see that relations in Eq. (\ref{condition1}), as well
as the associated ones $a_j\gamma_{kl}+\gamma_{jk}\gamma_{jl}=0$,
correspond to single zeros, since $\gamma_{jk}\neq 0$. One finds
$(a_k a_l-\gamma^2_{kl})/(a_j a_l-\gamma^2_{jl})
=(a_k+\gamma_{jk}\gamma_{kl}/\gamma_{jl})/(a_j+\gamma_{jk}\gamma_{jl}/\gamma_{kl})
=(\gamma_{kl}/\gamma_{jl})^2$.

\begin{figure}[t]
\psfig{figure=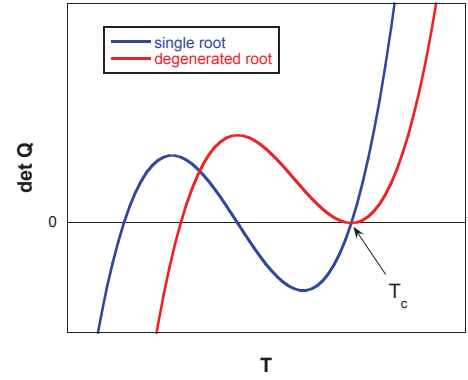,width=6cm} \caption{(color online). Schematic
temperature dependence of determinant of the coupling matrix of the
linearized GL equations (see Eqs.(\ref{LGL}) and (\ref{Tc})). The
doubly degenerated root at $T_{\rm c}$ is a necessary condition for
stable TRSB state, whereas the single-root case is associated with a
conventional, TRSR state similar to single-component systems. }
\end{figure}

The order parameters for $T\lesssim T_{\rm c}$ are given by

\vspace{-3mm}
\begin{equation}
\left[
\begin{array}{ccc}
  a_1+b_1|\psi_1|^2 & -\gamma_{12} & -\gamma_{13} \\
  -\gamma_{12} & a_2+b_2|\psi_2|^2  & -\gamma_{23} \\
  -\gamma_{13} & -\gamma_{23} & a_3+b_3|\psi_3|^2
\end{array}
\right]
\left[
\begin{array}{c}
  \psi_1 \\
  \psi_2 \\
  \psi_3
\end{array}
\right]
=\left[
\begin{array}{c}
  0 \\
  0 \\
  0
\end{array}
\right].
\label{NLGL}
\end{equation}

\noindent Putting $\psi_1$ real as always possible, the imaginary
parts in $\psi_2$ and $\psi_3$ should obey the relations

\vspace{-3mm}
\begin{equation}
\left[
\begin{array}{cc}
  a_2+b_2|\psi_2|^2  & -\gamma_{23} \\
  -\gamma_{23} & a_3+b_3|\psi_3|^2
\end{array}
\right]
\left[
\begin{array}{c}
  {\rm Im}(\psi_2) \\
  {\rm Im}(\psi_3)
\end{array}
\right]
=\left[
\begin{array}{c}
  0 \\
  0
\end{array}
\right].
 \label{op2}
\end{equation}

\noindent  Therefore, for complex order parameters one has
$a_3b_2|\psi_2|^2+a_2b_3|\psi_3|^2\simeq -a_2 a_3+\gamma^2_{23}$ up
to O$(t)$ with $t\equiv (T_{\rm c}-T)/T_{\rm c}$. In the same way,
one obtains two other similar relations, and thus

\vspace{-3mm}
\begin{equation}
\left[
\begin{array}{ccc}
  0 & a_3b_2 & a_2b_3 \\
  a_3b_1 & 0  & a_1b_3 \\
  a_2b_1 & a_1b_2 & 0
\end{array}
\right]
\left[
\begin{array}{c}
  |\psi_1|^2 \\
  |\psi_2|^2 \\
  |\psi_3|^2
\end{array}
\right]
=\left[
\begin{array}{c}
  -a_2a_3+\gamma^2_{23} \\
  -a_1a_3+\gamma^2_{13} \\
  -a_1a_2+\gamma^2_{12}
\end{array}
\right].
\label{op3}
\end{equation}

\noindent We then arrive at the following temperature dependence of
order parameters:

\vspace{-3mm}
\begin{equation}
|\psi_j|^2\simeq -(a_j+\gamma_{jk}\gamma_{jl}/\gamma_{kl})/b_j.
\label{abs-op}
\end{equation}

\noindent up to ${\rm O}(t)$.

 The single independent relation in
Eq.(\ref{LGL}), for example,
$a_1-\gamma_{12}\psi_2/\psi_1-\gamma_{13}\psi_3/\psi_1=0$, is then
equivalent to

\vspace{-3mm}
\begin{equation}
\frac{a_1}{\sqrt{b_1}}+\frac{a_2}{\sqrt{b_2}} e^{i\phi_{21}} +
\frac{a_3}{\sqrt{b_3}} e^{i\phi_{31}}=0
\end{equation}

\noindent for $T\lesssim T_{\rm c}$, where $\phi_{21}$ ($\phi_{31}$)
is the phase difference between $\psi_2$ ($\psi_3$) and $\psi_1$. It
becomes clear that the condition for a state of complex order
parameters to be stable is equivalent to that of a triangle formed
by three segments:

\vspace{-3mm}
\begin{equation}
\frac{a_j}{\sqrt{b_j}}+\frac{a_k}{\sqrt{b_k}}>\frac{a_l}{\sqrt{b_l}}
\label{condition2}
\end{equation}

\noindent for $T\lesssim T_{\rm c}$. A phase diagram is displayed in
Fig.~3.

The relations (\ref{condition1}) and (\ref{condition2}) formulate
the full condition for the stability of state with complex order
parameters, \textit{i.e.} the TRSB superconductivity. It is clear
that the special case with isotropic parameters discussed previously
\cite{Gorkov, Tesanovic, Yanagisawa} satisfies these conditions.

A phase transition at a lower temperature $T_{\rm tr}<T_{\rm c}$
from a TRSB state to a TRSR state is possible for appropriate
temperature dependence of parameters (see Fig.~3), where interesting
physics is expected. Although discussions in the GL scheme can be
pushed forward, we notice that to treat the two transitions
concretely one needs a more microscopic theory.

\begin{figure}[t]
\psfig{figure=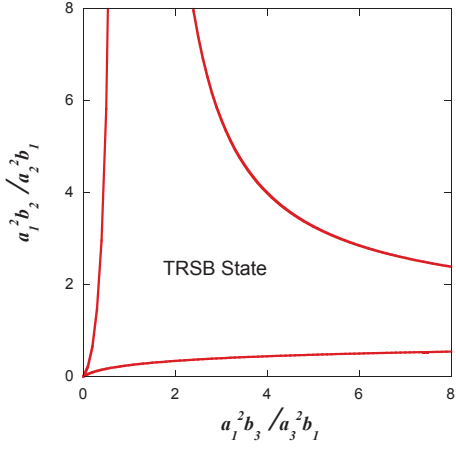,width=6cm} \caption{(color online). Phase
diagram for a three-component superconductor with stable TRSB
superconductivity at the central part and TRSR one at the corners.}
\end{figure}

Next we investigate the coherence length of the state of complex
order parameters. In order to demonstrate the novelty of this state,
we concentrate for a while on a system of isotropic parameters in
Eq.(\ref{GL}) except $m_1\equiv m$ and $m_2=m_3 \equiv m'$, the
simplest, but non-trivial case allowing analytic treatment. In the
bulk, the amplitude of the order parameter is $\psi_0\equiv
|\psi_j|=\sqrt{-(a+\gamma)/b}$, and the phase difference between any
of the two components is $2\pi/3$ \cite{Gorkov, Tesanovic,
Yanagisawa}. A local distortion in the first component
$\psi_1=(1+\delta_1)\psi_0$ causes distortions in the other two
components $\psi_2=(1+\delta_2)\psi_0\exp[i(2\pi/3+\delta_3)]$ and
$\psi_3=(1+\delta_2)\psi_0\exp[i(4\pi/3-\delta_3)]$, with
$\delta_j$'s real, as displayed in Fig.~4a. In the one-dimensional
case, the GL equation for the variation of $\psi_1$, for example, is
$a\psi_1+b\psi^2_0\psi_1-\gamma (\psi_2+\psi_3)
=(\hbar^2/2m)\partial^2\psi_1/\partial x^2$, which, with two other
similar equations, yield

\vspace{-3mm}
\begin{equation}
\begin{array}{cc}
  (a+3b\psi^2_0)\delta_1 + \gamma \delta_2 + \sqrt{3}\gamma \delta_3
    & = \frac{\hbar^2}{2m}\frac{\partial^2\delta_1}{\partial x^2},
    \\ \vspace{5mm}
  \frac{\gamma}{2}\delta_1 + (a+3b\psi^2_0+\frac{\gamma}{2})\delta_2
   -\frac{\sqrt{3}\gamma}{2}\delta_3
    & = \frac{\hbar^2}{2m'}\frac{\partial^2\delta_2}{\partial x^2},
    \\ \vspace{5mm}
  \frac{\sqrt{3}\gamma}{2}\delta_1 -\frac{\sqrt{3}\gamma}{2}\delta_2
  + (a+b\psi^2_0-\frac{\gamma}{2})\delta_3
    & = \frac{\hbar^2}{2m'}\frac{\partial^2\delta_3}{\partial x^2}.
\end{array}
\label{DFGL}
\end{equation}

\noindent The coherence length defined by $\delta_j=A_j\exp(-\sqrt{2}x/\xi)$
at large distance limit is thus determined as

\vspace{-3mm}
\begin{equation}
\frac{\hbar^2\xi^{-2}}{-(a+\gamma)m}=
\frac{3m'}{2m}\frac{2+\frac{m'}{m}\pm \sqrt{(\frac{m'}{m})^2-\frac{4m'}{3m}+\frac43}}{1+2m'/m}.
\label{coherenceL}
\end{equation}

\vspace{3mm} \noindent There are two divergent solutions, and the
corresponding characteristic modes are given by
$A_2/A_1=-R/[2(3-2R)]$ and $A_3/A_1=-\sqrt{3}(2-R)/[2(3-2R)]$, where
$R$ is defined by the right-hand side of Eq.(\ref{coherenceL}).
Mode-I associated with the larger solution, thus giving the
coherence length of the system, is specified by $R=1$, and
$A_2/A_1=-1/2$ and $A_3/A_1=\sqrt{3}/2$ when $m=m'$ (see Fig.~4b),
whereas mode-II by $R=2$, and $A_2/A_1=1$ and $A_3/A_1=0$ (see
Fig.~4c), with the distortion vectors form an equilateral triangle
in both cases. While mode-II at $m=m'$ is the conventional one
associated merely with variation of amplitude (see Fig.~4c), known
in single- and two-component cases, mode-I is the novel one in which
variations of amplitude and phase are coupled (see Fig.~4b),
specific to the TRSB state.

\begin{figure}[t]
\psfig{figure=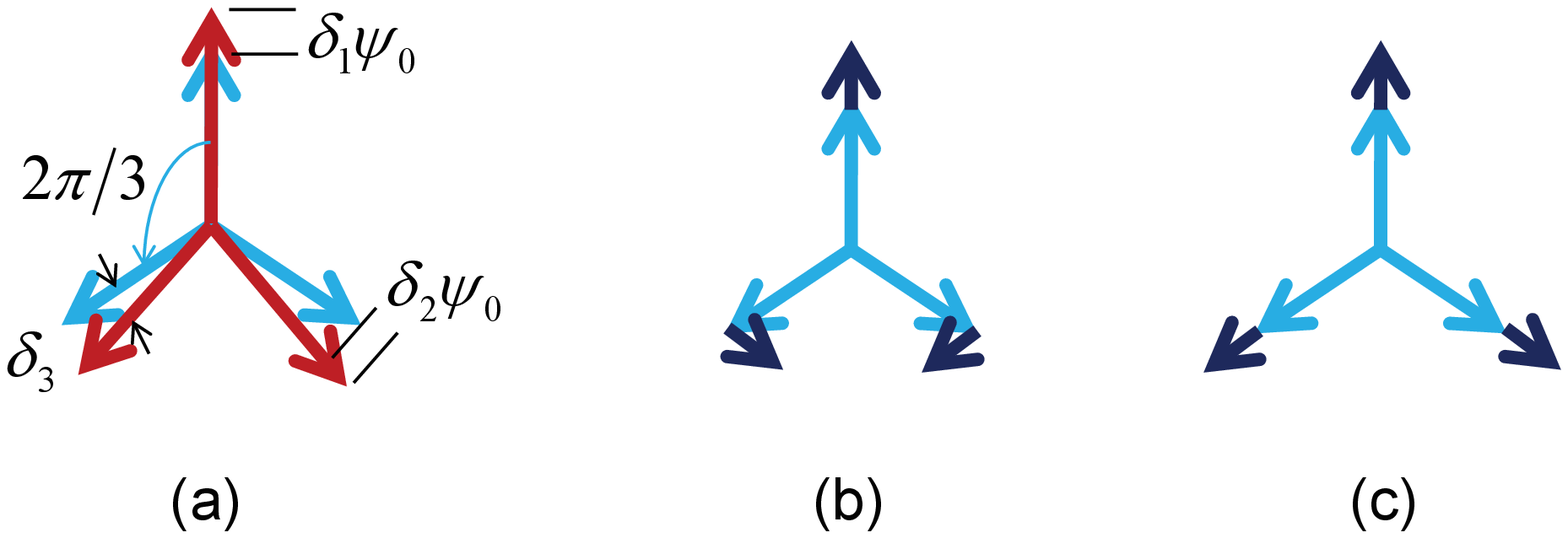,width=7cm} \caption{(color online).
Distortions in the complex order parameters in TRSB state of the
\textit{isotropic} system (see text). (\textbf{a}) Definitions of
distortions in order parameters. (\textbf{b}) and (\textbf{c})
Characteristic distortions of mode-I and model-II respectively, for
$m=m'$. }
\end{figure}

Other quantities for the present superconductivity are available
straightforwardly, such as the London penetration depth $\lambda$

\vspace{-3mm}
\begin{equation}
 \lambda^{-2}=\frac{4\pi(2e)^2}{c^2}(-1-\gamma_{12}\gamma_{13}/a_1\gamma_{23})
 \sum\limits_j a_j/b_jm_j,
 \label{penetrationD}
\end{equation}

\noindent the thermodynamic critical field $H_{\rm tc}$ from Eq.(\ref{abs-op})

\vspace{-3mm}
\begin{equation}
 \frac{H^2_{\rm tc}}{8\pi}=
 \frac{1}{2}(1+\gamma_{12}\gamma_{13}/a_1\gamma_{23})^2\sum\limits_j a^2_j/b_j,
 \label{Htc}
\end{equation}

\noindent and the nucleation field $H_{\rm n}$ (or $H_{\rm c2}$)

\vspace{-3mm}
\begin{equation}
 H^2_{\rm n}= \frac{\phi_0}{2\pi}\frac{{\rm det}{\bf Q}}{-(a_1a_2-\gamma^2_{12})a_3}
 \frac{1}{\sum\limits_j \hbar^2/2a_jm_j }.
 \label{Hn}
\end{equation}

It is clear that $H_{\rm tc}\neq \phi_0/2\sqrt{2}\pi\xi\lambda$, and
$H_{\rm n}\neq \phi_0/2\pi\xi^2$. Consequently, we are led to
conclude that this superconductivity cannot be
categorized by the GL number $\kappa\equiv \lambda/\xi$ into type I
or type II \cite{Tinkham}.

Because of the existence of two divergent solutions of
Eq.(\ref{coherenceL}), a vortex can exhibit different core sizes
\cite{Tinkham} for the three components even close to $T_{\rm c}$.
This may cause a long-range attractive and short-range repulsive
interaction between two vortices and thus exotic vortex states, a
possibility discussed previously in two-component case
\cite{Babaev}.

Finally let us investigate the Josephson current between two bulks
linked by a short, narrow constriction in otherwise continuous
superconducting material, as displayed in Fig.~1. The length of the
constriction is much shorter than the coherence length $l\ll \xi$.
Suppose that the two bulks exhibit opposite chiralities by chance in
a cooling process, with $\psi_{j{\rm L}}$ and $\psi_{j{\rm R}}$ the
wave functions at the left and right bulk respectively, where
$|\psi_{j{\rm L}}|=|\psi_{j{\rm R}}|\equiv \psi_{j0}$, $\psi_{1{\rm
R}}/\psi_{1{\rm L}}=e^{i\varphi}$, $\psi_{2{\rm L}}/\psi_{1{\rm
L}}=e^{i\phi_{21}}\psi_{20}/\psi_{10}$, $\psi_{3{\rm L}}/\psi_{1{\rm
L}}=e^{i\phi_{31}}\psi_{30}/\psi_{10}$, $\psi_{2{\rm R}}/\psi_{1{\rm
R}}=e^{i(2\pi-\phi_{21})}\psi_{20}/\psi_{10}$, and $\psi_{3{\rm
R}}/\psi_{1{\rm R}}=e^{i(2\pi-\phi_{31})}\psi_{30}/\psi_{10}$. The
order parameters on the bridge $\psi_j(x)\equiv f_j(x)\psi_{j{\rm
L}}$ are determined by $a_1 f_1+b_1\psi^2_{10}|f_1|^2
f_1-\gamma_{12}f_2\psi_{2{\rm L}}/\psi_{1{\rm L}}
 -\gamma_{13}f_3\psi_{3{\rm L}}/\psi_{1{\rm L}}
=(\hbar^2/2m_1)\partial^2 f_1/\partial x^2$, and two other similar
equations. Following the idea developed for conventional
single-component superconductors \cite{Larkin, Tinkham}, one gets the
linear solutions $f_1(x)=(1-x/l)+e^{i\varphi}x/l$,
$f_2(x)=(1-x/l)+e^{i(2\pi-2\phi_{21}+\varphi)} x/l$ and
$f_3(x)=(1-x/l)+e^{i(2\pi-2\phi_{31}+\varphi)} x/l$, with $x=0,l$ at
the left and right ends of the link, since $(\hbar^2/2m_j)\partial^2
f_j/\partial x^2={\rm O}(\psi^2_{j0})\propto \xi^{-2}\sim 0$ for
$l\ll \xi$.

The Josephson current between the two bulks is then \cite{Larkin,
Tinkham}

\begin{equation}
I=i_1\sin\varphi+i_2\sin(\varphi-2\phi_{21})
 +i_3\sin(\varphi-2\phi_{31}),
 \label{JoseI}
\end{equation}

\noindent with $i_j\equiv e\hbar^2\psi^2_{j0}/m_j$. We arrive at the
critical Josephson current

\vspace{-3mm}
\begin{equation}
 I_{\rm c}=
 \sqrt{\sum_j i^2_j +2\sum_{j<k}i_ji_k\cos2\phi_{jk}},
\label{criticalJoseI}
\end{equation}

\noindent which equals zero in the isotropic case. On the other
hand, when the two bulks share the same chirality, one has
\cite{Larkin,Tinkham} $ I=(i_1+i_2+i_3)\sin\varphi$ in contrast to
Eq.(\ref{JoseI}).  Therefore, because of the interference among the
components, the critical Josephson current between two bulk
superconductors of different chiralities is much smaller than the
one for same chirality.

In experiment based on the constriction structure shown
schematically in Fig.~1, the two bulk superconductors can show
either the same chirality or the opposite ones by chance in repeated
cooling processes. Then the standard measurement of critical
Josephson current will give two sets of values, quite different from
each other as discussed above, which gives a smoking gun evidence
for this novel superconductivity.

To finish we notice that although the GL approach is, in principle,
justified only close to the critical point, the phenomena revealed
in present are expected for the whole temperature regime.

\vspace{3mm}

\noindent \textbf{Acknowledgements} This work is upported by WPI
Initiative on Materials Nanoarchitectonics, MEXT of Japan, and
Grants-in-Aid for Scientific Research (No.22540377), JSPS, and
partially by CREST, JST. The authors thank M.~Tachiki, T.~Yanagisawa
and S.-Z.~Lin for discussions, and L.-H.~Wu for technical help.


\begin{thebibliography}{99}

\bibitem{BCS} J.~Bardeen, L.~N.~Cooper, J.~R.~Shrieffer,
\textit{Phys. Rev.} \textbf{106},
162 (1957); \textbf{108}, 1175 (1957).

\bibitem{Nambu} Y.~Nambu, \textit{Phys. Rev.} \textbf{117}, 648 (1960).

\bibitem{SigristUeda} M.~Sigrist, K.~Ueda,
\textit{Rev. Mod. Phys.} \textbf{63}, 239 (1991).

\bibitem{Maeno} G.~M.~Luke \textit{et al.},
\textit{Nature} \textbf{394}, 558 (1998).

\bibitem{Gorkov} D.~F.~Agterberg, V.~Barzykin, L.~P.~Gor'kov,
 \textit{Phys. Rev. B} \textbf{60}, 14868 (1999).

\bibitem{Tesanovic} V.~Stanev, S.~Te\v{s}anovi\'{c},
\textit{Phys. Rev. B} \textbf{81}, 134522 (2010).

\bibitem{Yanagisawa} Y.~Tanaka, T.~Yanagisawa,
\textit{J. Phys. Soc. Jpn.} \textbf{79}, 114706 (2010).

\bibitem{Hosono} Y.~Kamihara, T.~Watanabe, M.~Hirano, H.~Hosono,
\textit{J. Am. Chem. Soc.} \textbf{130}, 3296 (2008);
for a review see K.~Ishida, Y.~Nakai, H.~Hosono,
 \textit{J. Phys. Soc. Jpn.} \textbf{78}, 062001 (2009).

\bibitem{NgNagaosa} T.~K.~Ng, N.~Nagaosa,
 \textit{Europhys. Lett.} \textbf{87}, 17 (2009).

\bibitem{SCZhang} W.~C.~Lee, S.~C.~Zhang, C.~Wu,
\textit{Phys. Rev. Lett.} \textbf{102}, 217002 (2009)

\bibitem{Ota} Y.~Ota, M.~Machida, T.~Koyama, H.~Aoki,
 \textit{Phys. Rev. B} \textbf{83}, 060507(R) (2011).

\bibitem{Tinkham} M.~Tinkham, \textit{Introduction to Superconductivity},
McGraw-Hill, Inc. Second Edition (1996).

\bibitem{Gurevich} A.~Gurevich, \textit{Physica C} \textbf{456}, 160 (2007).

\bibitem{ArfkenWeber} G.~B.~Arfken, H.~J.~Weber, \textit{Mathermatical Methods
for Physicists}, Elsevier Academic Press, Sixth Edition (2005).

\bibitem{Kondo}  J.~Kondo, \textit{Prog. Theor. Phys.} \textbf{29}, 1 (1963).

\bibitem{note1} It is easy to see that the root of Eq.(\ref{Tc})
cannot be triply degenerated since the summation of the eigenvalues
of the matrix ${\bf Q}$ is $\sum_{j=1,2,3}a_j>0$.

\bibitem{Babaev} E.~Babaev, M.~Speight,
\textit{Phys. Rev. B} \textbf{72}, 180502(R) (2005).

\bibitem{Larkin} L.~G.~Aslamozov, A.~I.~Larkin,
\textit{JETP Lett.} \textbf{9}, 87 (1969).

\end{thebibliography}
\end{document}